\documentclass[11pt]{article}

\def\half{\frac{1}{2}}
\usepackage{axocolor}
\usepackage{a4wide}
\usepackage{latexsym}
\usepackage{epsfig}
\usepackage{amsmath}
\usepackage{amssymb}

\def\beq{\begin{equation}}
\def\eeq{\end{equation}}
\def\barr{\begin{array}}
\def\earr{\end{array}}
\def\dis{\displaystyle}
\def\U{{\cal U}}

\newlength{\dinwidth} \newlength{\dinmargin}
\begin{document}

\newcommand{ \slashchar }[1]{\setbox0=\hbox{$#1$}   
   \dimen0=\wd0                                     
   \setbox1=\hbox{/} \dimen1=\wd1                   
   \ifdim\dimen0>\dimen1                            
      \rlap{\hbox to \dimen0{\hfil/\hfil}}          
      #1                                            
   \else                                            
      \rlap{\hbox to \dimen1{\hfil$#1$\hfil}}       
      /                                             
   \fi}                                             %
\def\delslash{\slashchar{\partial}}
\def\pslash{\slashchar{p}}
\def\qslash{\slashchar{q}}
\def\aslash{\slashchar{a}}
\def\xslash{\slashchar{x}}
\def\xtslash{\slashchar{\tilde x}}

\thispagestyle{empty}

\begin{flushright}
IMSc/2008/03/03
\end{flushright}

\vspace{1.5cm}

\begin{center}
 {\Large\bf Fermionic un-particles, gauge interactions and
the $\beta$ function}\\[1cm] 
  {\sc  Rahul Basu$^a$, Debajyoti Choudhury$^b$, H. S. Mani$^a$\\
  \vspace{1.5cm}
  $^a$ {\it The Institute of Mathematical Sciences, CIT Campus,
  Taramani, Chennai 600 113, India}\\[0.4cm]
  $^b${\it Department of Physics and Astrophysics, University of Delhi, 
  Delhi 110 007, India} \\
}
\end{center}

\vspace{2cm}


\begin{abstract}

\noindent{The dynamics of fermionic unparticles is developed 
from first principles. It is shown that any unparticle, whether 
fermionic or bosonic, can be recast in terms of a canonically  
quantized field, but with non-local interaction terms.
We further develop a possible gauge theory for fermionic unparticles.  
Computing the consequent contribution of un-fermions  to the
$\beta$ function of the theory, it is shown that this can be viewed 
as the sum of two contributions, one fermion-like and the 
other scalar-like. However, if full conformal invariance is 
imposed, the latter vanishes identically. 
We discuss the consequences thereof as well as some general phenomenological
issues.}
\end{abstract}

\vspace*{\fill}

\newpage
\reversemarginpar

\section{Introduction}
\label{sec:introduction}
The idea of elementary fields with non--integer scaling dimension as
proposed in \cite{Georgi:2007ek} and `deconstructed' in
\cite{Stephanov:2007ry} has received considerable attention. Such fields can,
perhaps, be best motivated in a nontrivial scale invariant theory with
an infrared fixed point (examples being afforded by a vector-like
non-abelian gauge theory with a large number of massless fermions as
studied by Banks and Zaks (BZ)~\cite{Banks:1981nn}, or certain
nonlinear sigma models~\cite{Braaten:1985is}). Manifesting itself in
the existence of asymptotic states that are not particle-like (in the
sense of not having a well-defined mass, but rather a continuous mass
spectrum), such a field has been termed an
``unparticle''~\cite{Georgi:2007ek}.  In an interacting theory, the
aforementioned non-integer scaling dimension (though it has been shown
in \cite{Stephanov:2007ry} that the unparticle can be looked upon as an 
infinite  ladder
of ordinary particles in the limit of a vanishing energy gap) leads to
curious effects in its propagation as well as its kinematics. Such
effects are illustrated if the unparticle is allowed to couple to
standard model (SM) currents through an effective interaction and this
has given rise to a whole body of work that have
looked at various phenomenological implications of the existence of
such particles.

Direct phenomenological issues aside, unparticles are unusual, to say
the least~\cite{Georgi:2007si}. As we shall demonstrate below, the
unparticle field can be redefined to yield a canonical massless field
(thus preserving the signature scale invariance of this sector). The
price to be paid is the transmutation of any (local) interactions with
normal particles into non-local ones.  In other words, unparticles
provide us with a concrete realisation of the structure of non-local
field theories. Thus, even as a pure field theoretical problem, it is
interesting to study the behavior of unparticles in their interactions
with other matter and gauge particle.

Even though phenomenological consequences have been discussed 
at length, some of the more basic theoretical issues 
have not found too much space in the literature. In particular, changes
in the dynamics of a theory (e.g. the Standard Model) due to the
presence of unparticles has not quite been carefully delineated. Since
this can have significant effects on the running of the coupling
constants, and hence to the overall behavior of the theory at different
length scales, it is important to understand the nature of the $\beta$
function in a theory with unparticles.  

There has been some discussion in the literature on the effect of
scalar unparticles on a gauge theory \cite{Liao:2007fv}. There has also been
some discussion regarding fermionic unparticles~\cite{Luo:2007bq}. However,
the propagator proposed in the latter set of papers do not go over to the
usual fermionic propagator when the dimension of the field is taken to
$3/2$. As a result, the conclusions of the papers are somewhat
suspect.

In order to clarify all these issues we start from a Lagrangian of
fermionic unparticles, gauge it with the usual prescription of
introducing a Wilson line and proceed to calculate the $\beta$ function
contribution coming from these unparticles. For this we use the result
of \cite{Terning:1991yt} which treats the general case of local and non local
Lagrangians. 

In a recent paper, Grinstein et al. \cite{Grinstein:2008qk} have pointed out
certain important issues regarding the compatibility of the
propagators for unparticles used in the literature with conformal
field theory (CFT). We use in this paper, a propagator invariant under
scale transformations, Lorentz transformations, and translations but
not under special conformal transformations.  We thereafter comment on
the effect that special conformal transformations have on the
result. In order to clarify these somewhat subtle matters which have
been insufficiently dealt with in the literature, we explicitly show
in Appendix B the transformation of the fermionic 2-pt function under
the conformal group. The form of the 2 point function for fermions is
different depending on whether we demand invariance under the full
conformal group or not. It turns out that the $\beta$ function
obtained gets contribution from a spinor and a scalar part which would
lead to a different evolution of the coupling from that in a fully
conformally invariant theory where the $\beta$ function gets
contribution only from the spinor part as for a normal fermion. This
is essentially because in a fully conformally invariant theory we are
forced to choose a fermion propagator to be the same as a `normal'
fermion.

In Section II we demonstrate the derivation
of the fermionic unparticle propagator. In Section III we show that
a redefinition of the fermion fields allows us to rewrite the
Lagrangian in terms of a local and non local part and thereby calculate
the one loop fermionic corrections to the scalar propagator. 
Thereafter in Section IV we repeat this case for the corrections to the
gauge field propagator due to un-fermionic loops. In this case we
discuss the issue of gauge invariance and provide a prescription for
gauging the theory. In Section V, we calculate the full one loop two
point function for the non abelian gauge field due to un-fermion
and thence the correction to the $\beta$ function of the theory due to
un-fermionic modes. The consequences to phenomenology are discussed in a
subsequent Section VI.

Finally in the Appendices we tie up various loose ends left in the
paper. In particular we show that rescaling of the fields as done for
fermions could equally well have been done for scalars and we use conformal
field theory considerations to derive the form of the fermionic
propagator. 
\section{Fermionic unparticles}
\label{sec:ferm_unp}
Consider an unparticle fermion field $\Psi_\U(x)$ of dimension $d$.
The two-point correlator 
\beq \dis
{\cal T} \equiv  \int d^4 x \, e^{i \, P \cdot x} \;
    \langle0| T[\Psi_\U(0) \, \bar \Psi_\U(x)] | 0 \rangle
\eeq
should then be describable as a coherent superposition of a continuum 
of single particle propagators with an appropriate density of states. 
In other words, 
\beq
\barr{rcl}
{\cal T} & = & \dis
\int_0^\infty \frac{d M^2}{2 \, \pi} \; \rho(M^2) \; 
           \frac{\slashchar{P} +  M}{P^2 - M^2 + i \, \epsilon}
\\[2.5ex]
\rho(M^2) & = & \dis \frac{3}{2} \; A_{2 \, d / 3} \; (M^2)^{d - 5/2}
\\[2ex]
 A_d & \equiv & \dis \frac{16 \, \pi^{5/2}}{(2 \, \pi)^{2 \, d}} \; \;
                   \frac{\Gamma(d + 1 / 2)}{\Gamma(d - 1) \; \Gamma(2 \, d)}
\earr
    \label{derivation}
\eeq
where the exponent $(d - \, \frac{5}{2})$ is determined purely 
from dimensional arguments and $A_d$, as before,
is the phase space for the emission of $d$ massless 
particles\footnote{In general,  the time ordered product for the 
fermionic case may be defined by two independent spectral functions. 
Here, for simplicity, we shall choose them to be identical. 
We return to this point in Appendix B}. 
Note that $\rho(M^2)$ should have $A_{2 \, d / 3}$ (and not 
$A_{d}$ as in Ref.\cite{Luo:2007bq}) since the emission of 
a (composite) fermion of mass-dimension $d$ is equivalent to emission of 
$2 \, d / 3$ canonical fermions (which have mass dimension $3/2$). 
This has crucial 
implications as indicated below. On performing the integral explicitly, 
one obtains
\beq
\dis
{\cal T} = \frac{- 3}{4 \, \cos(\pi \, d)} \; 
A_{2 \, d / 3} \, (-P^2 - i \, \epsilon)^{d - 5/2} \;
\left[ \slashchar{P} -  \cot (\pi \, d) \; \sqrt{-P^2- i \, \epsilon} \, 
\right] \ .
      \label{propag_psi}
\eeq
This, then, defines the propagator for a fermionic unparticle.
Clearly, in the limit $d \to \frac{3}{2}^+$, the propagator 
reduces to that for a canonical massless fermion, as it rightly 
should. This is in marked contrast to the propagator as obtained in 
Ref.\cite{Luo:2007bq}. Furthermore, it should be easy to see that 
eq.(\ref{propag_psi}) satisfies the constraints imposed by 
scale invariance, namely, that the 
two point correlator of a field of dimension $d$ should go 
as $|x_1 - x_2|^{- 2 \, d}$ where $x_i$ denote the space-time 
coordinates.

It is, then, a straightforward exercise to construct an appropriate
Lagrangian density for a free fermionic unparticle, namely
\beq
{\cal L}_\Psi = \dis 
\frac{4 \, \cos(\pi \, d) \, \sin^2(\pi \, d)}{3 \; A_{2 \, d / 3}} \, 
\bar \Psi_\U(x) (\partial_\mu \, \partial^\mu)^{3/2- d} \;
\left[ i \delslash
       +  \cot (\pi \, d) \; \sqrt{\partial_\nu \, \partial^\nu}
\right]\;  \Psi_\U(x) \ .
     \label{lagr_psi}
\eeq
We believe that the above should be used for doing phenomenology with 
fermionic unparticles rather than that of Ref.~\cite{Luo:2007bq} as has been 
done in the literature. 

Certain important features should be noted at this stage: 
\begin{itemize}
\item The un-fermion, as described above, must have a scaling dimension 
$d \geq 3/2$. 
\item The propagator ${\cal T}$, while reducing 
to the canonical form for $d \to \frac{3}{2}^+$, is ill-defined 
for any $d = n + {1 \over 2}, \forall \, n > 1, n \in Z$. 
\item On the other hand, ${\cal T}$ is quite well-defined for 
$ d = { 3 \,  \over 2}$, belying the naive expectation that multi-particle 
cuts should not be expressible in terms of a single propagator.
\item The Lagrangian density of eq.(\ref{lagr_psi}) 
has the curious form of being the sum of two Lagrangians: 
a fermionic one and a scalar one, both unparticle-like. As a corollary,
so does ${\cal T}$. 
\end{itemize}
As would be obvious from the discussions 
in  the subsequent sections, the last point above 
is a very crucial one in the understanding of the fermionic 
unparticle described by eq.(\ref{lagr_psi}) and the penultimate point 
is but a consequence of this. 

\section{A redefinition of fields}
\label{sec:ferm_redefn}

It is instructive to reformulate 
the fermionic unparticle in terms of a different field. 
Effecting a field redefinition, viz. 
\beq
\barr{rcl}
\Psi_U(x) \to \chi(x) & \equiv & \dis 
\sqrt{\cal A} \, (\partial_\mu \, \partial^\mu)^{3/4- d/2} \, 
                            \Psi_\U(x)   
\\[2ex]
{\cal A} & \equiv & \dis 
\frac{4 \, \cos(\pi \, d) \; \sin^2(\pi \, d)}{3 \; A_{2 \, d / 3}}
\earr
\eeq
we may rewrite the Lagrangian density of eq.(\ref{lagr_psi}) as 
\beq
{\cal L}_\Psi = {\cal L}_\chi = \dis 
\bar \chi(x) \; 
\left[ i \delslash
       +  \cot (\pi \, d) \; \sqrt{\partial_\nu \, \partial^\nu}
\right]\; \chi(x) \ .
     \label{lagr_chi}
\eeq
Thus, $\chi(x)$ behaves like a canonically quantized field. 
The presence of the extra term may seem baffling at first, 
but a closer inspection shows that the operator
\[
\Sigma \equiv \cot (\pi \, d) \; 
(\partial_\mu \partial^\mu)^{1/2}
\]
is equivalent to a 
non-local mass term~\cite{Terning:1991yt} for $\chi(x)$. Indeed, the propagator 
can be expressed as 
\beq 
\barr{rcl}
S(p) & = & \dis \frac{i}{\pslash +  \Sigma_0(p)} 
       = i \, \sin^2(\pi \, d) \; 
              \frac{\pslash -  \Sigma_0(p)}{p^2 + i \, \epsilon}
\\[2ex]
\Sigma_0(p) & \equiv & \dis \cot (\pi \, d) \; (-p^2)^{1/2} \ .
\earr
\eeq
Since $\Psi_\U$ and $\chi$ represent  quite different entities, the   
two theories would result in very different predictions 
for asymptotic unfermion states. On the other hand, if the unfermions 
are to appear only as virtual states, the two theories clearly would be 
equivalent, as long as all its interactions are non-derivative in nature.
(Note that the above considerations are valid for redefinition of a
general field, scalar or spinor. The effects are similar to the above.
This has been explicitly shown in Appendix A).
To illustrate this, let us consider a Yukawa interaction between 
an un-fermion with a normal real scalar $\phi$ , viz.
\beq
      \barr{rcl}
      {\cal L} & = & \dis
      \frac{4 \, \cos(\pi \, d) \, \sin^2(\pi \, d)}{3 \; A_{2 \, d / 3}} \, 
      \bar \Psi_\U(x) (\partial_\mu \, \partial^\mu)^{3/2- d} \;
      \left[ i \delslash 
	+  \cot (\pi \, d) \; \sqrt{\partial_\nu \, \partial^\nu}
	\right]\;  \Psi_\U(x) 
      \\[2ex]
      & + & \dis y \; \bar \Psi_\U(x) \Psi_\U(x) \; \phi(x)
      + \frac{1}{2} (\partial_\mu \phi) \, (\partial^\mu \phi) 
      - \frac{m^2 \; \phi^2}{2} 
           \quad .
      \earr
           \label{yukawa_1}
\eeq
\begin{figure}[!h]
\vspace*{6ex}
 \begin{picture}(155,0)(-5.0,0)
\DashArrowLine(0,0)(70,0){5}{psOliveGreen}
    \Text(20, 15)[c]{$\phi(q)$}
\DashArrowLine(130,0)(200,0){5}{psOliveGreen}
    \Text(180, 15)[c]{$\phi(q)$}
\ArrowArc(100, 0)(30,0,180){psRed}
    \Text(100, 15)[c]{$\Psi_\U(p-q)$}
\ArrowArc(100, 0)(30,180,360){psRed}
    \Text(100, -15)[c]{$\Psi_\U(p)$}

\DashArrowLine(250,0)(320,0){5}{psOliveGreen}
    \Text(270, 15)[c]{$\phi(q)$}
\DashArrowLine(380,0)(450,0){5}{psOliveGreen}
    \Text(430, 15)[c]{$\phi(q)$}
\ArrowArc(350, 0)(30,0,180){psBlue}
    \Text(350, 15)[c]{$\chi(p-q)$}
\ArrowArc(350, 0)(30,180,360){psBlue}
    \Text(350, -15)[c]{$\chi(p)$}

\Text(100,-40)[c]{($a$)}
\Text(350,-40)[c]{($b$)}

\end{picture}
\vspace*{8ex}
\caption{\em The one-loop correction to a scalar propagator for a
scalar coupling to an un-fermion. Panels {\em (a)} and {\em (b)} 
correspond to the Lagrangians of eq.(\ref{yukawa_1}) and
eq.(\ref{yukawa_2}) respectively.}
    \label{fig:scal_prop}
\end{figure}
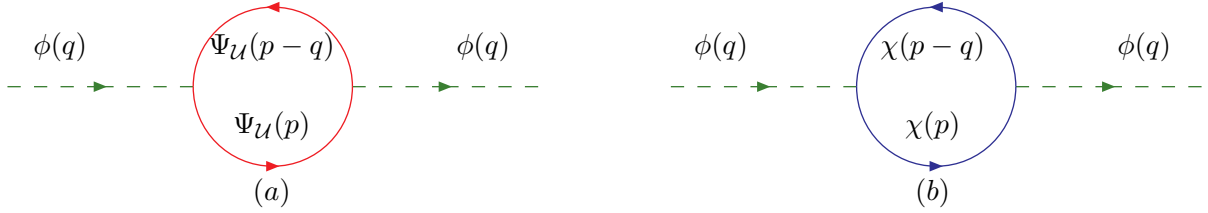
The one-loop correction to the $\phi$ propagator is then given by 
the diagram of Fig.\ref{fig:scal_prop}($a$) and amounts to 
\beq
\barr{rcl}
i \, \Pi(q^2) 
& = & \dis - Tr \; \int \frac{d^4 p}{(2 \, \pi)^4} \;
       (i \, y)^2 \; 
\frac{- 3 \; i}{4 \, \cos(\pi \, d)} \; 
A_{2 \, d / 3} \, (-p^2)^{d - 5/2} \;
\left[ \pslash
       -  \cot (\pi \, d) \; \sqrt{-p^2} 
\right]
\\[2ex]
&& \dis \hspace*{3em}
\frac{- 3 \; i}{4 \, \cos(\pi \, d)} \; 
A_{2 \, d / 3} \, \left[ - (p - q)^2\right]^{d - 5/2} \;
\left[ (\pslash - \qslash)
       -  \cot (\pi \, d) \; \sqrt{- (p - q)^2} 
\right]
\\[3ex]
& = & \dis
\frac{- 9 \, y^2 \,A_{2 \, d / 3}^2  }{4 \, \cos^2 (\pi \, d)} \; 
\int \frac{d^4 p}{(2 \, \pi)^4} \;
 \left[ p^2 \; (p - q)^2\right]^{d - 5/2} \;
\left[ p \cdot (p - q) 
       + \Sigma_0(p) \; \Sigma_0(p - q) \right] \ . 
\earr
     \label{phi_corr}
\eeq
Rewriting the theory in terms of the field $\chi(x)$, we have
      \beq
      \barr{rcl}
      {\cal L} & = & \dis
\bar \chi(x) \; 
\left[ i \delslash
       +  \cot (\pi \, d) \; \sqrt{\partial_\nu \, \partial^\nu}
\right]\; \chi(x) 
      + \frac{1}{2} (\partial_\mu \phi) \, (\partial^\mu \phi) 
      - \frac{m^2 \; \phi^2}{2}
      \\[2ex]
      & + & \dis 
     \frac{3 \, A_{2 \, d / 3} \; y }
          {4 \, \cos(\pi \, d) \,\sin^2 (\pi \, d) } \; 
 \; \left[ \left(\partial^2\right)^{d/2 - 3/4} \bar \chi(x) \right]
   \; \left[ \left(\partial^2\right)^{d/2 - 3/4} \chi(x) \right]\; \phi(x)
      \earr \ .
           \label{yukawa_2}
\eeq
The correction to the $\phi$ propagator is now given by 
Fig.\ref{fig:scal_prop}($b$) and is 
\[
\barr{rcl}
i \, \Pi(q^2) 
& = & \dis - Tr \; \int \frac{d^4 p}{(2 \, \pi)^4} \;
     \left[ \frac{3 \, A_{2 \, d / 3} \; y }
          {4 \, \cos(\pi \, d) \,\sin^2 (\pi \, d) }  \right]^2 \; 
 \left[ (-p^2)^{d/2 - 3/4} \; \left\{- (p- q)^2 \right\}^{d/2 - 3/4}\right]^2
\\[2ex]
& & \dis \hspace*{3em}
\left[ \pslash +  \Sigma(p)\right]^{-1} 
\; \left[ (\pslash - \qslash) +  \Sigma(p - q) \right]^{-1}  \ .
\earr
\]
This is easily seen to be the same as eq.(\ref{phi_corr}), thus vindicating 
our assertion that as long as the unfermion is to be treated only as 
virtual states, the $\Psi_\U$ and the $\chi$ theories would yield 
identical results.

Before we end this section, note that the scaling dimension $d$ 
appears only as a (periodic) parameter in the $\chi$ theory. For 
$d \to \frac{3}{2}^+$, it once again reduces to a canonical fermion 
as it must, while for integral values of $d$ it behaves purely as 
an un-scalar. 

\section{Gauging the un-fermion}
\label{sec:gauging}

Although most of the unparticle literature assumes unparticles to be 
gauge singlets, it is interesting to consider the possibility 
that these have gauge interactions~\cite{Cacciapaglia:2007jq,Liao:2007fv}. 
While it is 
quite conceivable that the unparticles may transform non-trivially 
under a gauge group orthogonal to the SM, nothing, in principle, 
forbid them from having nonzero SM quantum numbers too. 

The act of gauging the unfermion Lagrangian raises an important issue,
namely whether to gauge the $\Psi_\U$ theory or the $\chi$ theory. 
A ``minimal'' coupling for $\Psi_\U$ is not the same as that for 
$\chi$ and vice-versa. 
As is quite apparent, our previous assertion (of the two 
theories being equivalent when restricted to only virtual unfermions) 
does not hold any longer. This, though, is not unexpected
and is related to the derivative coupling typical of gauge interactions. 
A similar situation would be faced for ordinary (SM) particles as well, 
were it not for the fact that experiments point us to the right choice. 

In the absence of any phenomenological input, there is no {\em a priori} 
reason to prefer one theory over the other, and 
we must make a choice as to 
the ``right'' field to gauge. In this paper, we are guided by simplicity and 
choose to couple the field $\chi(x)$ to a nonabelian gauge field 
staying as close to minimal substitution as possible. Indeed, the 
formalism developed by Terning~\cite{Terning:1991yt} can be readily 
used in this context. As is expected, the presence of the 
$\sqrt{\partial_\nu \, \partial^\nu}$ term in the Lagrangian 
leads to an infinite number of interaction terms of the generic form 
$A^n \, \bar\chi \chi$. For our purpose, it suffices to consider 
only the three-point and four-point vertices. Defining 
\beq
\barr{rcl}
\dis
\Sigma_1(p;q) & \equiv & \dis 
\frac{\Sigma_0(p+q) - \Sigma_0(p)}{(p+q)^2 - p^2} \ ,
\\[3ex]
\Sigma_2(p;q_1, q_2) & \equiv & \dis
\frac{\Sigma_1(p; q_1 + q_2) - \Sigma_1(p; q_1)}
      {(p+q_1 + q_2)^2 - (p + q_1)^2}  \ , 
\earr
\eeq
the three point function is given by~\cite{Terning:1991yt}
\beq
\barr{rcl}
\Gamma^\mu(p, q, p+q) & = & \dis
       i \, g \, T_a \, \left[ \gamma_\mu + (2 \, p + q)_\mu \;
                        \Sigma_1(p ; q) \right]
\earr
\eeq
where $p$ ($p + q)$ is the momentum of the incoming (outgoing) unfermion field 
$\chi$ and $q$ is the momentum of the incoming gluon. Similarly, 
the 4-point function is seen to be
\beq
\barr{rcl}
\Gamma^{\mu\nu}(p, q_1, q_2, p+q_1+q_2) & = & \dis
i \, g^2 \, \Big[ \{T_a, T_b \}\, g^{\mu \nu} \; 
                   \Sigma_1(p ; q_1+q_2) 
\\[2ex]
&  & \dis \hspace*{1em} + 
T_a \, T_b \; 
(2 \, p + q_2)^\nu \; \left[2 \, (p + q_2) + q_1\right]^\mu
\; \Sigma_2(p; q_2, q_1)
\\[2ex]
& & \dis \hspace*{1em} +
T_b \, T_a \; 
(2 \, p + q_1)^\mu \; \left[2 \, (p + q_1) + q_2\right]^\nu
\; \Sigma_2(p ; q_1, q_2)
\Big]
\earr
\eeq
where $p$ ($p + q_1 + q_2)$ is the momentum of the incoming (outgoing) 
fermion field 
$\chi$ and $q_{1\mu}$ and $q_{2\nu}$ are the momenta of the incoming gluons
(with color indices $a$ and $b$ respectively). The Lorentz structure of 
the extra piece in $\Gamma^\mu$ and the first term in $\Gamma^{\mu\nu}$ 
are both reflective of a scalar-gauge field interaction. This is not 
unexpected as both these terms owe their origin to the scalar-like 
$\sqrt{\partial_\nu \, \partial^\nu}$ term in the Lagrangian. Indeed, the
non-local ``mass'' term can be thought of as a scalar inextricably intertwined 
with the fermion. Consequently, $\Sigma_0(p)$ contributes to the 
propagator, while the first differential $\Sigma_1(p;q)$ is the coefficient 
of the scalar-gluon coupling resultant from the first fermion-scalar-fermion
intermingling. The higher differentials ($\Sigma_2$ onwards) can be thought 
of similarly. 

Before we end this section, note that
\beq \dis
q_\mu \; \Gamma^\mu(p, q, p+ q)
= i \, g \, T_a \, 
       \left[ S^{-1}(p + q) - S^{-1}(p) \right]
\eeq
thus satisfying the Ward identity.

\section{The $\beta$ function}
\label{sec:betafn}

\begin{figure}[!h]
\vspace*{15ex}
 \begin{picture}(155,0)(-55.0,20)
\Gluon(0,0)(60,30){5}{7.5}{psOliveGreen}
    \Text(20, 28)[c]{$g^a_\mu(q)$}
\Gluon(120,0)(60,30){5}{7.5}{psOliveGreen}
    \Text(100, 29)[c]{$g^b_\nu(q)$}
\ArrowArc(60, 60)(30,0,360){psBlue}
    \Text(65, 75)[c]{$\chi(p)$}

\Gluon(150,50)(220,50){5}{7.5}{psOliveGreen}
    \Text(170, 65)[c]{$g^a_\mu(q)$}
\Gluon(280,50)(350,50){5}{7.5}{psOliveGreen}
    \Text(330, 65)[c]{$g^b_\nu(q)$}
\ArrowArc(250, 50)(30,0,180){psBlue}
    \Text(250, 65)[c]{$\chi(p)$}
\ArrowArc(250, 50)(30,180,360){psBlue}
    \Text(250, 35)[c]{$\chi(p+q)$}

\Text(50,-10)[c]{($a$)}
\Text(250,-10)[c]{($b$)}

\end{picture}
\vspace*{6ex}
\caption{\em The one-loop contributions to the vacuum polarization
from un-fermions.}
    \label{fig:vac_pol}
\end{figure}
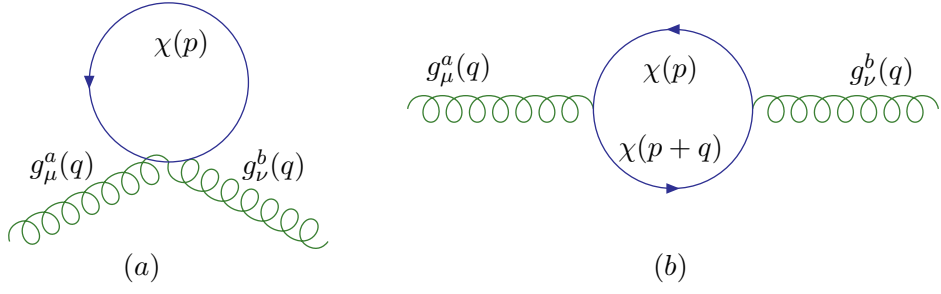

With the un-fermion having gauge interactions, the renormalization 
group evolution of the gauge coupling constant $g$ would receive 
contributions from unfermion loops. While each of vacuum polarization, 
un-fermion self-energy, unfermion gauge vertices (single gluon as well as
multiple gluons) as well as the gauge boson self-interactions receive 
corrections, the generalized Ward identities ensure that, as far as 
the $\beta$-function is concerned, one needs to consider only the 
additional contribution to the vacuum polarization. The corresponding
diagrams, to one-loop order, are displayed in Fig.\ref{fig:vac_pol}. 

The first diagram (involving the 4-pt vertex) gives 
\beq
\barr{rcl}
\dis 
i \, \Pi^{ab}_{\mu \nu}(q; {\rm 4pt}) 
& = & \dis - Tr \int \frac{d^4 p}{(2 \, \pi)^4} \; 
           \frac{i}{\pslash +  \Sigma_0(p)} \;
          i \, \Gamma_{\mu \nu}(p, q, -q, p)
\\[3ex]
& = & \dis \,g^2 \,  Tr \int \frac{d^4 p}{(2 \, \pi)^4} \; 
           \frac{\pslash -  \Sigma_0(p)}{p^2 - \Sigma_0^2(p)} \;
 \\[2ex]
& & \dis
\Bigg\{
\{T_a, T_b \}\, g_{\mu \nu} \; 
                   \Sigma_1(p ; 0) 
+
T_a \, T_b \; 
(2 \, p - q)_\nu \; (2 \, p - q)_\mu
\; \Sigma_2(p; -q, q)
\\[3ex]
& + & \dis 
T_b \, T_a \; 
(2 \, p + q)_\mu \; (2 \, p + q)_\nu
\; \Sigma_2(p ; q, -q)
\Bigg\} \ .
\earr
\eeq
Noting that $\Pi^{ab}_{\mu \nu}$ is invariant under 
$q \leftrightarrow -q$, the second piece in the integral 
is identical to the third. Thus, 
\[
\barr{rcl}
i \, \Pi^{ab}_{\mu \nu}(q ; 4pt) 
& = & \dis -2 \, g^2 \,  tr (T_a \, T_b) \; \sin^2 (\pi \, d ) \; 
Tr(1) \\[2ex]
& & \dis
 \int \frac{d^4 p}{(2 \, \pi)^4} \; 
           \frac{\Sigma_0(p)}{p^2  } \;
\Bigg\{g_{\mu \nu} \; \Sigma_1(p ; 0) 
 + (2 \, p + q)_\mu \; (2 \, p + q)_\nu \; \Sigma_2(p ; q, -q)
\Bigg\}
\earr
\]
where $\Sigma_1(p; 0)  =  \Sigma_0(p) / (2 \, p^2)$ and $Tr$ represents 
the Dirac algebra trace.  Some 
straightforward algebra results in 
\[
\barr{rcl}
i \, \Pi^{ab}_{\mu \nu}(q ; 4pt) 
& = & \dis -2 \, g^2 \,  tr (T_a \, T_b) \; \sin^2 (\pi \, d ) \; 
Tr(1) \\[2ex]
& & \dis
 \int \frac{d^4 p}{(2 \, \pi)^4} \; 
           \frac{1}{p^2  } \;
\Bigg\{ \frac{\cot^2(\pi \, d)}{2} \; 
\left[ \frac{(2 \, p + q)_\mu \; (2 \, p + q)_\nu }{2 \, p \cdot q + q^2}
\, -  g_{\mu \nu}  \right]
 \\[2ex]
& & \dis \hspace*{6em}
+ \Sigma_0(p) \; \Sigma_1(p;q) \;
\frac{(2 \, p + q)_\mu \; (2 \, p + q)_\nu }{2 \, p \cdot q + q^2} \Bigg\} \ .
\earr
\]
The first piece $[ \cdots ]$ gives a contribution 
identical to the {\em total} contribution from a canonically 
quantized scalar in the same representation~\cite{Liao:2007fv},
and we finally have
\beq
\barr{rcl}
i \, \Pi^{ab}_{\mu \nu}(q ; 4pt) 
& = & \dis - \; \frac{Tr(1) }{2} \; \cos^2(\pi \, d) \; 
         \left[i \, \Pi^{ab}_{\mu \nu}\right]_{\rm normal \; \, scalar}
 \\[2ex]
& - & \dis 
g^2 \,  tr (T_a \, T_b) \; \sin^2 (\pi \, d ) \; 
Tr(1) \;
 \int \frac{d^4 p}{(2 \, \pi)^4} \; \frac{ C^{(4)}_{\mu \nu}}{p^2 \, (p + q)^2}
\ ,
\\[3ex]
C^{(4)}_{\mu \nu} & \equiv & \dis
2 \, \Sigma_0(p) \; \Sigma_1(p;q) \; (p + q)^2 \; 
\frac{(2 \, p + q)_\mu \; (2 \, p + q)_\nu }{2 \, p \cdot q + q^2} \ .
\earr
   \label{4pt-diag}
\eeq
Note that the entire contribution is proportional to 
$\cos^2(\pi \, d)$ and vanishes identically for $d \to \frac{3}{2}^+$ 
as it should (for then the unfermion goes over to a canonical fermion).
The extra proportionality constant $[- Tr(1) / 2]$ ( $=2$ in 4 dimensions)
is reflective of the fact that this contribution emanates from 
a fermion (and hence the negative sign) which obviously has twice the 
degrees of freedom compared to a (canonically quantized) scalar 
in the same representation.

The contribution from the second diagram in Fig.\ref{fig:vac_pol} 
(involving the 3-pt vertices) is 
\beq
\barr{rcl}
\dis 
i \, \Pi^{ab}_{\mu \nu}(q; 3pt) 
& = & \dis - tr(T_a \, T_b) \; Tr \int \frac{d^4 p}{(2 \, \pi)^4} \; 
         \Gamma_\mu(p,q,p+q) \;  S(p) \;
         \Gamma_\nu(p+q,-q,p) \; S(p + q) \ .
\earr
\eeq
Using $ \Sigma_1(p+q ; -q) = \Sigma_1(p ; q)$,  this can be reduced to
\beq
\barr{rcl}
\dis 
i \, \Pi^{ab}_{\mu \nu}(q; 3pt) 
& = & \dis - g^2 \; tr(T_a \, T_b) \; \sin^4(\pi \, d) \;
\int \frac{d^4 p}{(2 \, \pi)^4} \; \left[p^2 \, (p + q)^2 \right]^{-1}
\\[2ex]
& & \dis \hspace*{1em}
      \Big\{ Tr [\gamma_\mu \, \pslash  \gamma_\nu \, (\pslash + \qslash)]
 +   Tr(1) \; C^{(3)}_{\mu \nu}\Big\} \ ,
\\[2ex]
C^{(3)}_{\mu \nu} 
& \equiv & 
(2 \, p + q)_\mu \; (2 \, p + q)_\nu \; \Sigma_1^2(p ; q) \, 
\left[ \Sigma_0(p) \, \Sigma_0(p+q) 
+ p  \cdot ( p +  q) 
\right]
\\[2ex]
&- & \dis 
   \left[ p_\mu \, (2 \, p + q)_\nu + (2 \, p + q)_\mu \; p_\nu \right]
\; \Sigma_1(p ; q) \, \Sigma_0(p+q) 
\\[2ex]
& - & \dis 
  \Sigma_1(p ; q) \, \Sigma_0(p) \; 
\left[ (2 \, p + q)_\mu \; (p + q)_\nu 
          + (2 \, p + q)_\nu \; (p + q)_\mu \right]
\\[2ex]
& + & \dis 
    \Sigma_0(p) \, \Sigma_0(p+q) \, g_{\mu \nu} \ .
\earr
\eeq
The first term in the integral, apart from the $\sin^4(\pi \, d)$
factor, is readily seen to be identical to the contribution of a
canonical fermion to the vacuum polarization\cite{Liao:2007fv}.  
Shifting $p \to -(p +
q)$, the second line in $C^{(3)}_{\mu \nu}$ can be seen to
be equivalent to the third line. Next, concentrating on the second
term of $C^{(3)}_{\mu \nu}$, we see
\[
\barr{l}
\dis 
\int \frac{d^4 p}{(2 \, \pi)^4} \; 
\frac{p  \cdot ( p +  q)}{p^2 \, (p + q)^2 } \; 
(2 \, p + q)_\mu \; (2 \, p + q)_\nu \; \Sigma_1^2(p ; q) 
\\[2ex]
\dis 
= - \int \frac{d^4 p}{(2 \, \pi)^4} \; 
\frac{p  \cdot ( p +  q)}{(2 \, p \cdot q + q^2)^2} \; 
(2 \, p + q)_\mu \; (2 \, p + q)_\nu \;
\left[\cot^2(\pi \, d) \; \left\{ \frac{1}{(p+q)^2} + \frac{1}{p^2} \right\}
+ \frac{           2 \, \Sigma_0(p + q) \, \Sigma_0(p)}
     {p^2 \, (p + q)^2 }
\right]
\earr
\]
and the first piece above can be seen to be equal to the second one by
shifting $p \to - (p + q)$.  Similarly, the first term of
$C^{(3)}_{\mu \nu}$ can be recast as
\[
\barr{l}
\dis
\hspace*{-2em}\int \frac{d^4 p}{(2 \, \pi)^4} \; 
    \frac{(2 \, p + q)_\mu \; (2 \, p + q)_\nu}{p^2 \, (p + q)^2} \;
\Sigma_1^2(p ; q) \, \Sigma_0(p) \, \Sigma_0(p+q) 
\\[2.ex]
 \dis = 
- \cot^2(\pi \, d) \; 
\int \frac{d^4 p}{(2 \, \pi)^4} \; 
    \frac{(2 \, p + q)_\mu \; (2 \, p + q)_\nu}{2 \, p \cdot q + q^2} \;
\Sigma_1(p ; q) \, 
             \left[ \frac{\Sigma_0(p)}{p^2} \,  -  \,
                   \frac{\Sigma_0(p+q) }{(p + q)^2} \right]
\\[2.ex]
 \dis = 
- 2 \, \cot^2(\pi \, d) \; 
\int \frac{d^4 p}{(2 \, \pi)^4} \; 
    \frac{(2 \, p + q)_\mu \; (2 \, p + q)_\nu}{2 \, p \cdot q + q^2} \;
\Sigma_1(p ; q) \, 
             \frac{\Sigma_0(p)}{p^2} \ .
\earr
\]
Thus, finally,
\beq
\barr{rcl}
\dis 
i \, \Pi^{ab}_{\mu \nu}(q; 3pt) 
& = & \dis \sin^4(\pi \, d) \; 
\left[i \, \Pi^{ab}_{\mu \nu}\right]_{\rm normal \; \, fermion}
\\[2ex]
& - & \dis 
g^2 \; tr(T_a \, T_b) \; \sin^4(\pi \, d) \;Tr(1) \; 
\int \frac{d^4 p}{(2 \, \pi)^4} \; \left[p^2 \, (p + q)^2 \right]^{-1}
\widetilde C^{(3)}_{\mu \nu}
\\[2ex]
\widetilde C^{(3)}_{\mu \nu} 
& \equiv & 
(2 \, p + q)_\mu \; (2 \, p + q)_\nu \; 
\left[ p  \cdot ( p +  q) \, \Sigma_1^2(p ; q) 
   - 2 \, \cot^2(\pi \, d) \, \Sigma_1(p ; q) \, \Sigma_0(p) \, (p + q)^2
\right]
\\[2ex]
&- & \dis 
   2 \, \left[ p_\mu \, (2 \, p + q)_\nu + (2 \, p + q)_\mu \; p_\nu \right]
\; \Sigma_1(p ; q) \, \Sigma_0(p+q) 
\\[2ex]
& + & \dis 
    \Sigma_0(p) \, \Sigma_0(p+q) \, g_{\mu \nu} \ . 
\earr
    \label{3pt-diag}
\eeq
The first term in $\widetilde C^{(3)}_{\mu \nu}$ has the same structure 
as $C^{(4)}_{\mu \nu}$ allowing us to make a crucial cancellation. 
In other words, adding eqs.(\ref{4pt-diag}\&\ref{3pt-diag}), 
we now have
\beq
\barr{rcl}
i \, \Pi^{ab}_{\mu \nu}(q)
& = & \dis - \; \frac{Tr(1) }{2} \; \cos^2(\pi \, d) \; 
         \left[ i \, \Pi^{ab}_{\mu \nu} \right]_{\rm normal \; \, scalar}
+
\sin^4(\pi \, d) \; 
\left[ i \, \Pi^{ab}_{\mu \nu} \right]_{\rm normal \; \, fermion}

\\[2ex]
& - & \dis 
2 \, g^2 \,  tr (T_a \, T_b) \; \sin^4 (\pi \, d ) \; 
Tr(1) \; \widehat\Pi_{\mu \nu}

\\[3ex]
\widehat \Pi_{\mu \nu} & \equiv & \dis
\int \frac{d^4 p}{(2 \, \pi)^4} \; \frac{1}{p^2} \; 
    \left[ G^{(1)}_{\mu \nu} + 2 \, \cot^2(\pi \, d) \, G^{(2)}_{\mu \nu} 
    \right]
\\[3ex]

G^{(1)}_{\mu \nu} & \equiv & \dis
\frac{\Sigma_0(p) \, \Sigma_0(p+q)}{(p+q)^2} \;
\Bigg\{ 
\frac{2 \, q  \cdot ( p +  q)}{(2 \, p \cdot q + q^2)^2} \;
(2 \, p + q)_\mu \; (2 \, p + q)_\nu 
\\[2.5ex]
& & \hspace*{6em} \dis  
+ 2 \, 
 \frac{\left[ p_\mu \, (2 \, p + q)_\nu + (2 \, p + q)_\mu \; p_\nu \right]}
      {2 \, p \cdot q + q^2} 
+ g_{\mu \nu}
\Bigg\}

\\[3ex]
G^{(2)}_{\mu \nu} & \equiv & \dis
\frac{ 1 } {(2 \, p \cdot q + q^2)} \; 
\Bigg\{ \left[ p_\mu \, (2 \, p + q)_\nu + (2 \, p + q)_\mu \; p_\nu \right]
 - p  \cdot q \, \frac{(2 \, p + q)_\mu \; (2 \, p + q)_\nu }{(2 \, p \cdot q + q^2)} \; 
\Bigg\} \ .
\earr
    \label{self_sum}
\eeq
It then remains to calculate only $\widehat\Pi_{\mu \nu}$. Rather 
than calculating it explicitly, note that it can always be expressed 
as
\[
\barr{rcl}
\widehat \Pi_{\mu \nu}(q) & = & \dis 
\frac{g_{\mu \nu} }{n - 1} \, \widehat \Pi_{\alpha \beta} \, 
          \left[ g^{\alpha \beta}
         - \,  \frac{q^\alpha \, q^\beta}{q^2}  
          \right] + 
    \frac{q_\mu \, q_\nu}{n - 1} \, \widehat \Pi_{\alpha \beta} \, 
          \left[ \frac{-g^{\alpha \beta}}{q^2} \; 
             + \frac{n}{q^4} \; q^\alpha \, q^\beta
          \right]
\earr
    \label{Pi_gaugeinv_proposition}
\]
where $n$ is the dimension of spacetime. Now,
\[
\int \frac{d^4 p}{(2 \, \pi)^4} \; q^\mu \, q^\nu \; 
\frac{1}{p^2} \, G^{(2)}_{\mu \nu} = 
\int \frac{d^4 p}{(2 \, \pi)^4} 
\frac{p \cdot q}{p^2}  =  0
\]
whereas 
\[
\barr{rcl}
\dis \int \frac{d^4 p}{(2 \, \pi)^4} \; q^\mu \, q^\nu \; 
\frac{1}{p^2} \, G^{(1)}_{\mu \nu} 
& = & \dis
3 \, \int \frac{d^4 p}{(2 \, \pi)^4} \; 
\frac{\Sigma_0(p) \, \Sigma_0(p+q)}{p^2 \; (p+q)^2} \;
q \cdot (2 \, p + q) 
= 0
\earr
\]
as the integrand is odd under the $p \to - (p + q)$ shift. In other 
words, $\widehat \Pi_{\mu \nu}$ is proven to be gauge invariant and 
it remains only to calculate $g^{\mu \nu} \, \widehat \Pi_{\mu \nu}$. 
Now,
\beq 
\barr{rcl}
\dis
g^{\mu \nu} \, G^{(1)}_{\mu \nu}
& = & \dis
\frac{\Sigma_0(p) \, \Sigma_0(p+q)}{(p+q)^2} \;
\Bigg\{ 
\frac{q^2 \; (2 \, p + q)^2}{(2 \, p \cdot q + q^2)^2} \;
+ n - 2 
\Bigg\}
\\[4ex]
g^{\mu \nu} \, G^{(2)}_{\mu \nu} & = & \dis
\frac{ 1 } {(2 \, p \cdot q + q^2)} \; 
\Bigg\{
\frac{ - p  \cdot q}{(2 \, p \cdot q + q^2)} \; 
(2 \, p + q)^2 
+ 2 p \cdot (2 \, p + q)
\Bigg\} \ .
\earr
\eeq
Defining
\beq
    y \equiv 2 \, p \cdot q + q^2 \ ,
\eeq
we may write 
\beq
\barr{rcl}
\dis g^{\mu \nu} \, \frac{1}{p^2} \, G^{(2)}_{\mu \nu} 
& = & \dis \; 
\frac{ 2 \, q^2 }{y^2} \; 
+ q^2 \; \frac{ p  \cdot q}{p^2 \, y^2} 
+  \frac{ 2 } {y}  \ .
\earr
\eeq
Since $y \to - y$ when $p \to - (p + q)$,  the integral 
of $y^{-1}$ vanishes identically, and we have
\beq 
g^{\mu \nu} \, \frac{1}{p^2} \, G^{(2)}_{\mu \nu} \longrightarrow 
 \frac{ 2 \, q^2 }{y^2} \; 
+ q^2 \; \frac{ p  \cdot q}{p^2 \, y^2}   \ . 
     \label{g2_final}
\eeq
To calculate the corresponding integral for $G^{(1)}_{\mu \nu}$, we first 
affect a Taylor expansion for $\Sigma_0(p + q)$, namely
\beq
\barr{rcl}
\Sigma_0(p + q) 
& = & \dis \Sigma_0(p) \; \left[ 1 + \frac{y}{2 \, p^2} 
                                   - \, \frac{y^2}{8 \, p^4}  + \cdots \right]
 \ .
\earr
\eeq
This is permissible since the $\beta$-function is only sensitive to 
ultraviolet physics. With this, 
\[
\barr{rcl}
\dis
g^{\mu \nu} \, \frac{1}{p^2} \, G^{(1)}_{\mu \nu}
& = & \dis
- \cot^2(\pi \, d) \;
\frac{1}{(p+q)^2} \;
\Bigg\{ 
\frac{q^2 \; (2 \, p + q)^2}{y^2} \;
+ n - 2 
\Bigg\} \;
\left[ 1 + \frac{y}{2 \, p^2} 
                                   - \, \frac{y^2}{8 \, p^4}  + \cdots \right]
\ .
\earr
\]
A bit of algebra, along with repeated use of the $p \to - (p + q)$ 
shift to throw away terms leads us to 
\beq
\dis
g^{\mu \nu} \, \frac{1}{p^2} \, G^{(1)}_{\mu \nu}
 \longrightarrow 
- \cot^2(\pi \, d)  \;
\Bigg\{\frac{4 \, q^2}{y^2}  \, - \, \frac{q^4 }{p^2 \, y^2} 
+  \frac{q^2 }{ 2 \, p^2 \, (p+q)^2} 
+ (n - 2 ) \, 
   \left[ \frac{1}{p^2}\,   - \, \frac{q^2}{8 \, p^4}  \right]
\Bigg\} + \cdots 
     \label{g1_final}
\eeq
where the ellipses contain terms ${\cal O}(p^{-5})$ which only 
lead to finite contributions. And, finally,
\beq
\barr{l}
\hspace*{-2em}
\dis \frac{g^{\mu \nu}}{p^2} \;
\left[ G^{(1)}_{\mu \nu} + 2 \, \cot^2(\pi \, d) \, G^{(2)}_{\mu \nu} 
    \right]
= 
\cot^2(\pi \, d) \; \Bigg\{ \frac{q^2 }{p^2 \, y} 
- \, \frac{q^2 }{ 2 \, p^2 \, (p+q)^2} 
- (n - 2 ) \, 
   \left[ \frac{1}{p^2}\,   - \, \frac{q^2}{8 \, p^4}  \right] \Bigg\} \ .
\earr
\eeq
Once again, the use of $p \to - (p + q)$ leads to a vanishing 
contribution from the first two terms, leaving only the last two. 
However, note that, in dimensional regularization, each of these 
vanish identically! Thus, $\widehat \Pi_{\mu \nu} = 0$ and
\beq
i \, \Pi^{ab}_{\mu \nu}(q)
= \sin^4(\pi \, d) \; 
\left[ i \, \Pi^{ab}_{\mu \nu} \right]_{\rm normal \; \, fermion}
- \; \frac{Tr(1) }{2} \; \cos^2(\pi \, d) \; 
         \left[ i \, \Pi^{ab}_{\mu \nu} \right]_{\rm normal \; \, scalar} \ .
\eeq

Several points are apparent at this stage:
\begin{itemize}
\item We do expect a form like this as 
       \begin{itemize}
           \item the $\chi$ propagator is essentially a sum of a local 
                 fermion propagator and a non-local scalar propagator;
           \item the $\bar \chi \chi g$ vertex is sum of a local 
                 fermion-fermion-gluon and a (non-local) scalar-scalar-gluon
		 vertex; 
           \item the coefficients are essentially a measure 
	         of the scalar and vector fractions.
       \end{itemize}
\item For $d = 3/2$, it does reduce to the canonical fermion case.
\item The relative sign between the fermion-like and scalar-like contributions
      was expected:
      \begin{itemize}
	\item the contribution from the $\slashchar{p}$ part of the propagator 
	      should behave like the 
	      canonical fermion contribution;
	\item note that the canonical scalar contribution has a sign opposite 
	      to that due to a canonical fermion. 
      \end{itemize}
\item The factor $Tr(1) / 2  \; (= 2)$ is a reminder of the fact that 
      each fermion mode is now associated with a scalar.

\item Note that this contribution to the 
      $\beta$-function can now have either sign depending on 
      the value of $d$. This is not unexpected for a change in $d$ results 
      in varying the relative content of fermion and scalar in $\chi$.

\item The above relation holds for both abelian and non-abelian theories. 
      With a proper choice of gauge representations for the unparticles
      [$SU(3)$, $SU(2)$ 
      as well as hypercharges], one can easily ensure gauge-coupling 
      unification if one were so inclined to.
\end{itemize}

\section{Phenomenology} 
At this stage, it is necessary to consider phenomenological
consequences of having an unparticle with nontrivial SM quantum
numbers. For example, what would the rate of production of a charged
unparticle in $e^+ e^-$ collisions be and what would its signature be?
Had we gauged the original field $\Psi_\U$, this question,
perhaps, would be best answered in the deconstructed
framework~\cite{Stephanov:2007ry}. With the unparticle now
corresponding to an infinite tower (with a vanishing mass gap) of
particles each coupled identically (and non-minimally) to the photon,
an $e^+ e^-$ collision would now result in pair production of
particles that are identical in electromagnetic properties. In other
words, we would be faced with pair production characterized by a
continuum in the mass of the particle, which, in the presence of
magnetic fields, would translate to a continuum in track curvature.
(It should be remembered that the unparticle cannot decay.) 
The absence of such signals would indicate a very small coupling
constant/charge, the nature of which continues to be an unresolved problem in
unparticles.

Gauging the Lagrangian for the $\chi$-field changes the situation
considerably.  As Sec.\ref{sec:gauging} shows, we now have a single
massless field (albeit with a non-local propagator) and with non-local
interactions with the electromagnetic field. The definition of
asymptotic states is straightforward and does not need
deconstruction. A non-zero charge for $\chi$ would result in copious
production of $\chi$-pairs. 
With the effective (non-local) mass of 
the each $\chi$ essentially given by $E_{\rm c.m.} / 2$, the electromagnetic 
radiation is an interesting
issue in itself. However, the absence of any
such nonstandard signal in an electromagnetic calorimeter 
would indicate the nonexistence of a light charged unparticle. 

One resolution of this problem would be to dispense with the
masslessness of the $\chi$ field. 
With the unparticle coupling to the SM, the lack of scale
invariance in the latter 
would also manifest itself in the unparticle sector as a
result of quantum corrections. In other words, once such effects are
taken into account in constructing the effective theory for
unparticles, the integral in eq.(\ref{derivation}) would be
characterized by an infrared cutoff $\Lambda_{IR}$, the magnitude of
which would be determined by the nature of the interaction with the SM.
Apart from a resultant complicated propagator, 
the $\chi$ field would still be characterized 
by a minimal mass $\sim {\cal O}(\Lambda_{IR})$ as well as 
a non-local term in its propagator. Phenomenologically, still,
the aforementioned problems would persist unless $\Lambda_{IR}$ is
sufficiently large.

What if the unparticle is an $SU(2) \otimes U(1)$ singlet, but is
coloured?  In an $e^+ e^-$ collider, $\chi$ production would now occur
only as a higher order process, say as $e^+ e^- \to q \bar q \chi \bar
\chi, \, g \chi \bar \chi$. The latter being a loop process, the
ensuing contribution to the 3-jet rate is relatively small and well
below the detection level. The 4-jet cross section, on the other
hand, could be significant and worth comparing with
the LEP observations of the slight excess
that was seen at ALEPH.  At a hadronic collider, though, the $q \bar q, gg \to 
\chi \bar \chi$ rate is quite large. Note, though, that the $\chi$'s would 
give rise to un-hadrons/un-jet which could contain 
various stable exotic bound states. 
Non-observation of such states would once again require that  $\Lambda_{IR}$ is
significantly large.  A large $\Lambda_{IR}$ essentially 
means that the SM $\beta$-functions would stay unaffected until 
$Q^2 \approx \Lambda^2_{IR}$ and the unparticle contribution(s) would manifest 
themselves only thereafter. 

Finally, a simple way of avoiding  
the above mentioned phenomenological constraints, is to make 
the unparticle gauge group different from the SM ones. Any interaction 
with the SM particles would then proceed strictly through a messenger sector 
as originally proposed by Georgi. The un-gauge is then a part of a shadow 
(mirror) world with its own dynamics and phenomenology (including candidates 
for dark matter etc.) and the exercise undertaken in this paper being 
applicable only towards an understanding of the same.

\section{Summary}
In this paper, we start by expressing the two point correlator of a
fermionic unparticle as a coherent superposition of a continuum of
single particle propagators convoluted with the appropriate density of
states. The corresponding propagator has the correct limiting
properties (quite unlike the correlator considered so far in the
literature). This, then, allows us to write the correct free-field 
Lagrangian for such states. 

Unparticles have, so far, been described by a non-canonical kinetic
energy term with an effective local Lagrangian describing interactions
with Standard Model fields. We show here, though, that a field
redefinition allows us to rewrite the theory in terms of a canonically
quantized field but with a nonlocal interaction Lagrangian. This holds
for all unparticles, whether bosnic or fermionic. That the two related
theories would give identical results for cases where the unparticle
appears only as virtual states is easy to appreciate. We demonstrate
this through explicit calculations as well. Of course, for asymptotic
unparticle states, the two theories do give different results, as they
indeed should.

We next consider a field theory consisting of fermionic unparticles
coupled to a gauge (gluon) field. Using standard methods of field
theory, we have calculated the contribution to the $\beta$ function of
the theory coming from this fermionic unparticle--gauge particle
coupling. To carry out this procedure, it becomes necessary to decide
whether to impose full conformal invariance on the theory, or only
demand scale invariance. As we show in this paper, the consequences
are quite different. Demanding full conformal invariance produces for
us a contribution which is exactly the same as that for an ordinary
fermion, modulo an overall constant.  On the other hand, demanding
only scale invariance, allows us a different form of the fermion
propagator and consequently a contribution to the $\beta$ function
which is a combination of a scalar and a fermionic part.

Since the dynamics of the theory depends on the $\beta$ function, it is
important to understand the issue of conformal invariance {\it vis a
vis} a theory of unparticles. The differences arising from these issues
would have consequences for phenomenology in any theory that contains
couplings of unparticles to other particles. Some of these issues have
been touched upon in the paper. 

\section*{Acknowledgements}
The authors thank Romesh Kaul for many illuminating discussions.  DC
acknowledges support from the Department of Science and Technology,
India under project number SR/S2/RFHEP-05/2006 and HSM acknowledges
support from the Department of Atomic Energy, India.

\appendix 

\section{Appendix A}
\setcounter{equation}{0}
\label{sec:appendix-scalar_unp}
In this appendix we show that the rescaling done for fermionic
unparticles could equally well have been done for scalar unparticles. 
Starting from the Lagrangian \footnote{Note that for a $d$ dimensional 
field $\phi$ for $d \neq 1$ the interaction term is not conformally invariant,
though it can be made so by changing the power of the $\phi^*\phi$ term. 
However, this is not our primary concern in this Appendix. Similarly, the 
kinetic energy term would, in principle, need an $A_d$ term.} 
   \[
     {\cal L} = \phi^* \, (\partial_\mu \partial^\mu)^{2-d} \, \phi 
     + \frac{\lambda}{\Lambda^{4(d-1)}} (\phi^*\phi)^2 
   \]
we redefine, by introducing the field $\omega$
   \[
     \phi \to \omega \equiv (\partial_\mu \partial^\mu)^{(1 - d) / 2} \, \phi
   \]
The Lagrangian, in terms of $\omega$, becomes
$$
{\cal L} = \omega^* \partial_\mu \partial^\mu \omega + 
\frac{\lambda}{\Lambda^{4(d-1)}} \left[\omega^*
(\partial_\mu\partial^\mu)^{d-1} \omega \right]^2.
$$
Clearly, this rescaling results in a non-local interaction term which
can be dealt along the lines of Ref. \cite{Terning:1991yt}
\section{Appendix B}
\setcounter{equation}{0}
\label{sec:cft}
In this Appendix, we discuss the connection between Conformal Invariance
(CI) and the form of the fermion propagator. Although most of the
analysis in this section is well known, we place it here for the record
and in the context of fermionic unparticles.

The conformal algebra in 4 dimensions has 15 generators, 
$P_\mu$ (4 translations), $M_{\alpha\beta}$ (6 Lorentz transformations),
$D$ (1 scale transformation) and $K_\mu$ (4 special conformal transformations)
 with the following algebra:
$$
i[P_\mu,P_\nu]=0, i[M_{\alpha\beta},P^\mu]=\delta^\mu_{[\alpha}P_{\beta]},
i[M_{\alpha\beta},M^{\mu\nu}]= \delta^{[\mu}_{[\alpha}M_{\beta]}^{\ \ \nu]}
$$
and
$$
i[D,P_\mu]=P_\mu, i[D,K_\mu]=-K_\mu, i[D,M_{\alpha,\beta}]=0,
$$
$$
i[K_\mu,K_\nu]=0, i[M_{\alpha\beta},K^\mu]=\delta^\mu_{[\alpha}K_{\beta]},
i[P_\mu,K_\nu]= 2g_{\mu\nu}D - 2 M_{\mu\nu}
$$
The basic tranformations under translations, Lorentz transoformations,
scaling and special conformal transformations are as follows:\\
\noindent {\bf I. Translations}:\\
$ x^\mu\rightarrow x^{\prime\mu}=
x^\mu+\epsilon$
under which a generic field transforms as 
\begin{equation}
\phi(x)\rightarrow \phi^\prime(x^\prime)=\phi(x). 
\end{equation}
Thus
\begin{eqnarray}
\phi^\prime(x)= \phi(x-\epsilon)&=& (1-\epsilon.\partial)\phi(x) \nonumber \\
&=& \phi(x)-i\epsilon[P_\mu,\phi(x)] \ .
\end{eqnarray}

\noindent {\bf II. Lorentz tranformations}:\\
$x^\mu\rightarrow x^{\prime\mu}=
\Lambda^\mu_{\ \nu} x^\nu\simeq (\delta_\nu^\mu+\theta^\mu_{\ \nu})x^\nu $
under which a generic field transforms as
\begin{equation}
\phi(x)\rightarrow \phi^\prime(x^\prime)=M(\theta)\phi(x) \ .
\end{equation}
Thus, 
\begin{eqnarray}
\phi^\prime(x^\prime )&=& M(\theta) \, \phi(x) 
= (1+ \theta^{\mu\nu}M_{\mu\nu}) \, \phi(x) \ ,
\end{eqnarray}
where $M_{\mu\nu}$ is the spin part of the Lorentz generator, which for
scalar field $\phi(x)$, vector field $A^\alpha$ and spinor field $\psi(x)$
has the property
$$
M_{\mu\nu}\phi(x)=0, \quad 
M_{\mu\nu}A^\alpha(x)\equiv 
 \delta^\alpha_{[\mu}\delta^\beta_{\nu]}A_\beta=\delta^\alpha_{[\mu}A_{\nu]}
$$
and
$$
M_{\mu\nu}\psi(x) = \frac{1}{4}[\gamma_\mu,\gamma_\nu]\psi(x) \ .
$$

\noindent{\bf III. Scale Transformations}:\\
\begin{equation}
\phi(x) \rightarrow \phi^\prime(x^\prime)= e^{d\lambda}\phi(x) \simeq 
(1+d\lambda)\phi(x)
\end{equation}

\noindent{\bf IV. Special Conformal Transformation}:\\
\begin{equation}
x^\mu \rightarrow x^{\prime\mu}\equiv g(a)x^\mu = \frac{x^\mu+a^\mu x^2}
{1+2 \, a \cdot x + a^2 \, x^2}\simeq (1 - 2 \, a \cdot x) \, x^\mu 
 + x^2 \, a^\mu \ .
\end{equation}
Thus, for a scalar field $\phi(x)$ of mass dimension $d$, 
\begin{equation}
\phi(x)\rightarrow \phi^\prime(x^\prime)=[1 + 2 d \, a \cdot x  
      + 2  \, a^\mu \, x^\nu \, M_{\mu\nu}]\phi(x) \ .
\end{equation}

We are now equipped to consider the invariance properties of the various 
two point functions. Since in this paper we are concerned with the
fermion propagator, we will discuss only this case. The cases of the
scalar and vector propagators can be similarly calculated.

For a fermion field of mass dimension $d$ 

\beq
\barr{rcl}
\psi^\prime(x^\prime)&=& \dis 
  (1+2 d \, a \cdot x + 2 \, a^\mu x^\nu \, M_{\mu\nu})\psi(x) \\[1ex] 
	&=& (1+2da \cdot x + \frac{1}{2}[\aslash, \xslash])\psi(x) 
\\[1.5ex]
	{\bar \psi}^\prime  (x^\prime) &=& {\bar \psi}(x)(1+2 d \, 
	a \cdot x-\frac{1}{2} [\aslash, \xslash]) \ ,
\earr
\eeq
leading to 
\begin{eqnarray}
\label{prime_prop}
\langle {\psi^\prime(0) \, \bar \psi}^\prime(x^\prime) \rangle 
&=& (1+2d \, a \cdot x)\langle {\psi(0) \, \bar \psi}(x) \rangle
 - \frac{1}{2} \, 
     \langle { \psi(0) \, \bar \psi}(x)\rangle \, [\aslash,\xslash] \ .
\end{eqnarray}
Now, from scaling arguments, the two-point function has the form  
\begin{equation}
	\langle \psi(0) \, \bar \psi(x)\rangle  =
	\frac{A \, \xslash}{(x^2)^{d+\half}} + 
	\frac{B}{(x^2)^d}
\end{equation}
where $A$ and $B$ are dimensionless quantities, presumably dependent on $d$.

Therefore, using this form on the LHS and RHS of eq.(\ref{prime_prop})
\begin{equation}
	\frac{A\xslash^\prime}{(x^{\prime\,2})^{d+\half}}
	+ \frac{B}{(x^{\prime\,2})^d}
	= (1+2\, d\, a\cdot x)\left( \frac{A\xslash}{(x^2)^{d+\half}}
	+ \frac{B}{(x^2)^d}\right)-\half\left(\frac{A\xslash}{(x^2)^{d+\half}}
	+ \frac{B}{(x^2)^d}\right) [\aslash,\xslash] \ .
\end{equation}
Using $x^{\prime\,2}= x^2(1-2\, a\cdot x)$, we get for the LHS, expanding
to linear power in $a$
\begin{eqnarray}
&&	\frac{A[(1-2\, a\cdot x)\xslash +x^2\aslash]}{(x^2)^{d+\half}} 
	(1-2\, a\cdot x)^{-d-\half}+ \frac{B}{(x^2)^d}(1-2\, a\cdot
	x)^{-d}  \nonumber \\
&&	= \frac{A[1+(2\, d-1)\, a\cdot x]\xslash+x^2}{(x^2)^{d+\half}}
	+\frac{B}{(x^2)^d}(1+2\, d\, a \cdot x) \ .
\end{eqnarray}
On the other hand, the RHS of eq.(\ref{prime_prop}) reduces to 
\begin{equation}
	\frac{A[1+(2\, d-1)a\cdot x]\xslash+x^2}{(x^2)^{d+\half}}
	+ \frac{B}{(x^2)^d}\left(1+2\, d\, a\cdot
	x-\half[\aslash,\xslash]\right) \ .
\end{equation}

Thus for the Green's function to be conformally invariant, we would need 
$B=0$. On the other hand, if we dispense with the requirement of full conformal
invariance and demand only scale invariance (along with invariance under
Lorentz transformation and translations) we are allowed to keep both the
terms.



\begin{thebibliography}{10}
\bibitem{Georgi:2007ek}
  H.~Georgi,
  Phys.\ Rev.\ Lett.\  {\bf 98}, 221601 (2007)
  [arXiv:hep-ph/0703260].

\bibitem{Stephanov:2007ry}
  M.~A.~Stephanov,
  Phys.\ Rev.\  D {\bf 76}, 035008 (2007)
  [arXiv:0705.3049 [hep-ph]].

\bibitem{Banks:1981nn}
  T.~Banks and A.~Zaks,
  Nucl.\ Phys.\  B {\bf 196}, 189 (1982).

\bibitem{Braaten:1985is}
  E.~Braaten, T.L.~Curtright and C.K.~Zachos,
  Nucl.\ Phys.\  B {\bf 260}, 630 (1985).


\bibitem{Georgi:2007si}
  H.~Georgi,
  arXiv:0704.2457 .

\bibitem{Liao:2007fv}
  Y.~Liao,
  arXiv:0708.3327 [hep-ph].

\bibitem{Luo:2007bq}
  M.~Luo and G.~Zhu,
  Phys.\ Lett.\  B {\bf 659}, 341 (2008)
  [arXiv:0704.3532 [hep-ph]].


\bibitem{Terning:1991yt}
  J.~Terning,
  Phys.\ Rev.\  D {\bf 44}, 887 (1991).

\bibitem{Grinstein:2008qk}
  B.~Grinstein, K.~Intriligator and I.~Z.~Rothstein,
  arXiv:0801.1140 [hep-ph].


\bibitem{Gross:1970tb}
  D.~J.~Gross and J.~Wess,
  Phys.\ Rev.\  D {\bf 2}, 753 (1970).

\bibitem{Cacciapaglia:2007jq}
  G.~Cacciapaglia, G.~Marandella and J.~Terning,
  JHEP {\bf 0801}, 070 (2008)
  [arXiv:0708.0005 [hep-ph]].



\end{thebibliography}
\end{document}